\documentclass[aps,prl,twocolumn,superscriptaddress,10pt,showpacs]{revtex4-1}

\usepackage[utf8x]{inputenc}
\usepackage{graphicx}
\usepackage{amsmath,amssymb}
\usepackage{hyperref}
\usepackage{datetime}
\usepackage{color}

\newcommand{\ket}[1]{\left|#1\right\rangle}

\newcommand{\bk}[0]{\mathbf k}
\newcommand{\bp}[0]{\mathbf p}
\newcommand{\bb}[0]{\mathbf b}
\newcommand{\be}[0]{\mathbf e}

\newcommand{\br}[0]{\mathbf r}

\newcommand{\bra}[1]{\left\langle #1\right|}
\newcommand{\braket}[2]{\left\langle #1|#2\right\rangle}

\newcommand{\spl}[1]{\begin{align}\begin{split} #1 \end{split} \end{align}}

\newcommand{\vvec}[2]{\begin{pmatrix} #1 \\ #2 \end{pmatrix}}

\begin{document}

\title{Artificial graphene with tunable interactions}

\author{Thomas Uehlinger}
\affiliation{Institute for Quantum Electronics, ETH Zurich, 8093 Zurich, Switzerland}

\author{Gregor Jotzu}
\affiliation{Institute for Quantum Electronics, ETH Zurich, 8093 Zurich, Switzerland}

\author{Michael Messer}
\affiliation{Institute for Quantum Electronics, ETH Zurich, 8093 Zurich, Switzerland}

\author{Daniel\,\,Greif}
\email[]{greif@phys.ethz.ch}
\affiliation{Institute for Quantum Electronics, ETH Zurich, 8093 Zurich, Switzerland}

\author{Walter Hofstetter}
\affiliation{Institut f\"ur Theoretische Physik, Johann Wolfgang Goethe-Universit\"at, 60438 Frankfurt/Main, Germany}

\author{Ulf Bissbort}
\affiliation{Institut f\"ur Theoretische Physik, Johann Wolfgang Goethe-Universit\"at, 60438 Frankfurt/Main, Germany}
\affiliation{Singapore University of Technology and Design, 138682 Singapore}

\author{Tilman Esslinger}
\affiliation{Institute for Quantum Electronics, ETH Zurich, 8093 Zurich, Switzerland}

\date{\today}

\begin{abstract}
We create an artificial graphene system with tunable interactions and study the crossover from metallic to Mott insulating regimes, both in isolated and coupled two-dimensional honeycomb layers. 
The artificial graphene consists of a two-component spin mixture of an ultracold atomic Fermi gas loaded into a hexagonal optical lattice.
For strong repulsive interactions we observe a suppression of double occupancy and measure a gapped excitation spectrum. 
We present a quantitative comparison between our measurements and theory, making use of a novel numerical method to obtain Wannier functions for complex lattice structures. 
Extending our studies to time-resolved measurements, we investigate the equilibration of the double occupancy as a function of lattice loading time.
\end{abstract}

\pacs{
  05.30.Fk, 
  03.75.Ss, 
  67.85.Lm, 
  71.10.Fd, 
  71.30.+h, 
  73.22.Pr  
}

\maketitle

The engineering of systems that share their key properties with graphene~\cite{CastroNeto2009}, like Dirac fermions and a hexagonal structure, is gaining interest in an increasing number of disciplines in physics~\cite{Polini2013}. 
The artificial structures are created by confining photons in hexagonal lattices~\cite{Peleg2007,Kuhl2010}, by nanopatterning of ultra-high-mobility two-dimensional electron gases~\cite{Singha2011}, by scanning probe methods to assemble molecules on metal surfaces~\cite{Gomes2012} and by trapping of ultracold atoms in optical lattices~\cite{SoltanPanahi2011, tarruell_creating_2012}. 
The motivation for engineering graphene-like band structures is to explore regimes that are not, or not yet, accessible to research with graphene or similar materials. 
The artificial systems provide new avenues to topological~\cite{Haldane1988} and quantum spin Hall insulators~\cite{Kane2005, Guinea2010}, as well as to intriguing strongly correlated phases~\cite{meng_quantum_2010}. 
To understand the role of interactions in solids with complex lattice structures~\cite{Kotov2012}, ultracold fermionic atoms in optical lattices are particularly promising, as the inter-particle interactions and kinetic energy can be tuned~\cite{lewenstein_ultracold_2007, esslinger_fermi-hubbard_2010}, allowing for the realization of density and magnetic ordering~\cite{Jordens2008,Schneider2008,Greif2013}. 
In this Letter, we present and analyze a cold atoms based implementation of artificial graphene, where both the interaction and kinetic energy are tunable over a broad range.
For strong interactions we realize a 2D Mott insulator with ultracold fermions. 

To obtain a quantum degenerate Fermi gas we adhere to the procedure described in previous work~\cite{Jordens2008}. 
A balanced spin mixture of $^{40}\mathrm{K}$ atoms in the  $m_F=-9/2$ and $-7/2$ magnetic sublevels of the $F=9/2$ hyperfine manifold is evaporatively cooled in a crossed beam optical dipole trap to 15(2)\% of the Fermi temperature.
We prepare Fermi gases with total atom numbers between $N=25\times10^3$ and $300\times10^3$, with 10\% systematic uncertainty~\cite{jordens_quantitative_2010}.
We either set the scattering length to $86(2) a_0$ using a Feshbach resonance or transfer to an $m_F=(-9/2,-5/2)$ mixture, where we access more repulsive interactions in the range of $a=242(1) a_0$ to $632(12) a_0$ (the Bohr radius is denoted with $a_0$).

\begin{figure}[tb]
    \includegraphics{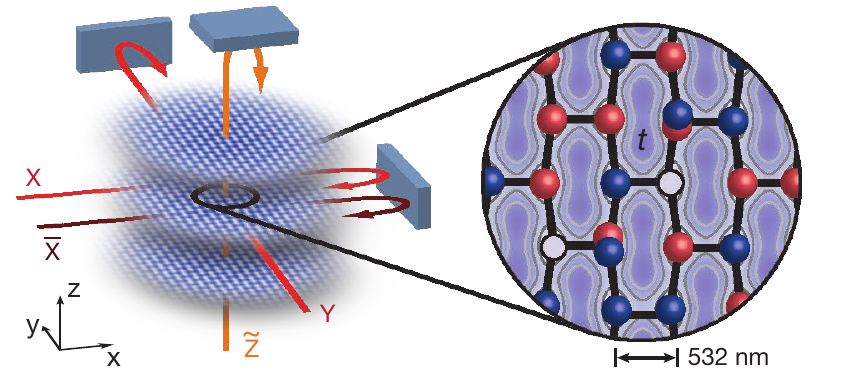}
    \caption{Experimental setup used to create the artificial graphene system. Independent 2D layers of honeycomb geometry are realized using a tunable-geometry optical lattice. A sketch of the tunneling structure within the layers is shown on the right. A repulsively interacting two-component spin mixture of fermionic $^{40}\mathrm{K}$ atoms (red and blue spheres) is loaded into the lattice. Gravity points along $y$.
	}\label{fig1}
\end{figure}

We then load the atoms into a tunable-geometry optical lattice~\cite{tarruell_creating_2012,Lim2012, uehlinger_double_2013} operating at a wavelength of $\lambda=1064\,\mathrm{nm}$ and consisting of three non-interfering, orthogonal standing-wave laser beams $\overline{X}$, $Y$ and $\widetilde{Z}$. An additional beam $X$ co-propagates with $\overline{X}$ and interferes with $Y$, see Fig.~\ref{fig1}. 
This gives rise to the potential
\begin{eqnarray} V(x,y,z) & = & -V_{\overline{X}}\cos^2(k_L
x+\theta/2)-V_{X} \cos^2(k_L
x)\nonumber\\
&&-V_{Y} \cos^2(k_L y) -V_{\widetilde{Z}} \cos^2(k_L z) \nonumber\\
&&-2\alpha \sqrt{V_{X}V_{Y}}\cos(k_L x)\cos(k_L
y)\cos\varphi , \label{eqlattice}
\end{eqnarray}
with $k_L=2\pi/\lambda$, visibility $\alpha=0.90(5)$, $\varphi=0.00(3)\pi$, and $\theta= 1.000(1)\pi$.
The final lattice depths in units of the recoil energy are $V_{{\overline{X}},X,Y,\widetilde{Z}}/E_R=$ $[14.0(4),0.79(2),$ $6.45(20),30(1)]$, unless explicitly stated otherwise. 
All beams are ramped up simultaneously to their final intensities within $200\,\mathrm{ms}$.
The resulting potential contains several independent 2D honeycomb layers with an inter-layer tunneling rate below $2\,\mathrm{Hz}$.
For the combined external confining potential of the dipole trap and the lattice laser beams we measure harmonic trapping frequencies of $\omega_{x,y,z}/2\pi = [86(2), 122(1), 57(1)]\,\mathrm{Hz}$. 

We characterize the state of our system by measuring the fraction of atoms on doubly occupied sites $D$~\cite{Jordens2008, jordens_quantitative_2010}.
To determine $D$, the tunneling is suppressed by switching off $V_X$ in roughly $5\,\mathrm{\mu s}$ and ramping up $V_{\overline{X}}$ and $V_Y$ linearly to a depth of $30\,\mathrm{E_R}$ within $500\,\mathrm{\mu s}$.
We then perform interaction dependent radio-frequency spectroscopy to obtain $D$~\cite{Jordens2008}. 
Both the independently determined offset in $D$ of $2.2(3)\%$ due to an imperfect initial spin mixture as well as the calibrated detection efficiency of $89(2)\%$ for double occupancies are taken into account~\cite{Greif2013}. 

\begin{figure}[tb]
    \includegraphics{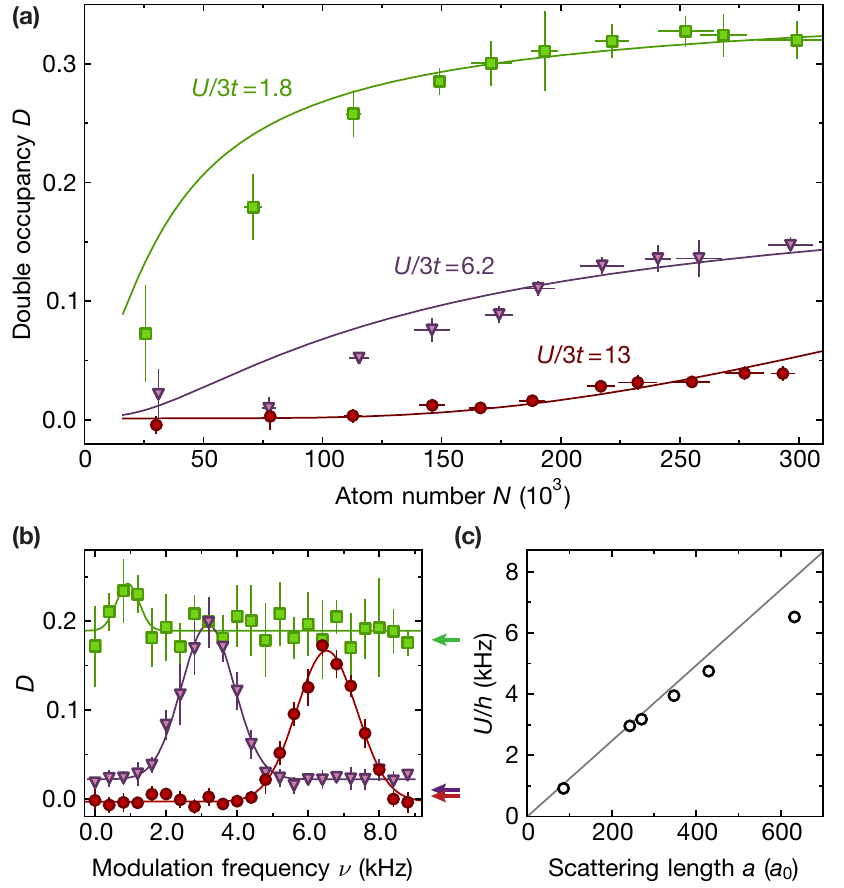}
    \caption{Observing the metal to Mott insulator crossover in artificial graphene. 
    {\bf (a)} The measured double occupancy~$D$ versus atom number $N$ for three different interaction strengths $U/3t$. 
    For strong interactions an incompressible Mott insulating core forms leading to a strong suppression of $D$.
    Solid lines are theory predictions based on a high temperature series expansion.
    {\bf (b)} Excitation spectrum obtained by measuring $D$ after sinusoidal modulation of the lattice depth $V_Y$ for the same interaction strengths as above.
    The solid lines are gaussian fits to the spectra.
    Arrows show the reference value without modulation.
    {\bf (c)} Comparison of the extracted Hubbard parameters $U$ with those obtained from a calculation of the Wannier functions in the honeycomb lattice.
    Errorbars in $D$ and $N$ show the standard deviation of 5 measurements. In panel (c), the uncertainty in $a$ and the fit error for the peak positions are smaller than the displayed data points.
    Data for additional interactions can be found in~\cite{supplementary}.
    Negative values of D are caused by the subtraction of an independently measured offset. 
    }\label{fig2}
\end{figure}

In the experiment we tune interactions from weakly ($U/3t=1.8(3)$) to strongly repulsive ($U/3t=13(1)$) and measure the double occupancy~$D$ as a function of the atom number $N$ in the lattice, see Fig.~\ref{fig2}(a). 
For weak interactions, the system is in a metallic state which is compressible, as signaled by an initial strong increase of $D$~\cite{Scarola2009}.
Here, creating more double occupancies requires less energy than placing additional atoms in the outside region of the harmonic trap where the potential energy is large.
For high atom numbers $D$ saturates as the system enters a band insulating state.
When increasing interactions, an incompressible Mott insulating state forms in the center of the trapped system. 
Therefore, $D$ is strongly suppressed and does not increase as more atoms are added to the system.
Only for the highest atom numbers the chemical potential becomes comparable to the on-site interaction, allowing for the creation of double occupancies~\cite{Jordens2008}.

A quantitative comparison of our results with a microscopic theory is made possible by describing our system by the Fermi-Hubbard Hamiltonian
\begin{equation}
    \hat H   =  -t\sum_{\langle ij\rangle,\sigma}(\hat c^\dagger_{i\sigma}\hat c_{j\sigma}+\mathrm{h.c.}) 
    + U\sum_i \hat n_{i\uparrow}\hat n_{i\downarrow} + \sum_{i,\sigma} V_i \hat n_{i\sigma}\;,
\end{equation}
where $\hat c^\dagger_{i\sigma}$ and $\hat c_{i\sigma}$ denote the fermionic creation and annihilation operators for the two spin states $\sigma\in\{\uparrow,\downarrow\}$ and $\langle ij\rangle$ denotes nearest neighbors.
The energy of the harmonic trap is $V_i$ and $\hat n_{i\sigma}=\hat c^\dagger_{i\sigma} \hat c_{i\sigma}$ is the density operator on site $i$. 
The determination of the on-site interaction energy $U$ and the nearest-neighbor tunneling matrix element $t$ requires an accurate calculation of the Wannier states, which is challenging for complex lattice structures such as used in our experiment.
To date the Marzari-Vanderbilt scheme \cite{Marzari1997, Mostofi2008685}, which numerically minimizes the Wannier functions' spatial variance, has become well established in the solid state community and has recently also been used for optical lattices \cite{Ibanez-Azpiroz2013,Walters2013,Ibanez-Azpiroz2013-2}. 
However, for complex lattice structures, this minimization may get stuck in local minima and becomes numerically expensive, requiring lattice-specific adaptations.
Instead, our numerical method (see \cite{supplementary}) is based on the alternative definition of Wannier states as eigenstates of band-projection operators \cite{Kivelson1982}, which we show to be a very suitable starting point for a numerical procedure.
The projection operator onto a subset of bands $\mathcal A$ can be written as $\mathcal P_{\mathcal A}=\sum_{\alpha\in \mathcal A, \, \bk} \ket{\bk,\alpha} \bra{\bk,\alpha}$, where $\ket{\bk,\alpha}$ is the 2D Bloch state with quasi-momentum $\bk$ in band $\alpha$ obtained from a standard band structure calculation of the lattice potential.
The Wannier states are then given by the simultaneous eigenstates of the two operators $R_j=\mathcal P_{\mathcal A} (\bb_j \cdot \hat \br) \mathcal P_{\mathcal A}$ with $\hat \br=(\hat x, \hat y)^T$ being real-space position operators, the reciprocal lattice vectors $\bb_j$ and $j=1,2$.
The calculation of the matrix elements $\bra{\bk,\alpha} R_j \ket{\bk', \alpha'}$ via real-space integration can be performed analytically \cite{supplementary}, reducing to a discrete summation of terms, which can be efficiently computed. 
In general $\mathcal A$ contains as many bands as there are sites per unit cell, i.e. two for the honeycomb lattice. 
The tunneling between nearest-neighbors and the interaction energy $U$ is subsequently determined in the usual way from Wannier function overlap integrals \cite{Jaksch2005}.
Our method is extendible to inhomogeneous systems, generic to all dimensions and allows proving the conjecture that the Wannier functions can be chosen to be real~\cite{supplementary}.
The method also holds for unit cells not symmetric under spatial reflection and if the Wannier states are no longer the Fourier transform of Bloch states.

We validate the qualitative interpretation of the data in Fig.~\ref{fig2}(a) using a high-temperature series expansion up to second order of the grand canonical partition function~\cite{Oitmaa2006} to determine the expected $D$.
For the calculation we use a nearest-neighbor tunneling of $t/h=172(20)\,\mathrm{Hz}$ within the layers ($h$ is Planck's constant) and separately measured on-site interaction energies $U/h=[0.92(12),3.18(2),6.52(3)]\,\mathrm{kHz}$ at the chosen scattering lengths $a=[86(2),$ $270(1),632(12)] a_0$.
The model assumes a connectivity of 3 within the 2D planes and no inter-layer tunneling, as well as a globally thermalized cloud.
Both finite temperature and the harmonic trap are taken into account.
We obtain overall good agreement with theory when allowing for the entropy per atom in the lattice $s=S/N$ as a fit parameter \cite{jordens_quantitative_2010}.
For the three interactions, the fitted entropies of $s=[2.1,2.7,1.7] k_\mathrm{B}$ are comparable to $s_\mathrm{in}=1.5(2)\,k_\mathrm{B}$ and $s_\mathrm{out}=2.5(1)\,k_\mathrm{B}$ measured in the dipole trap before loading and after reversing the loading procedure ($k_\mathrm{B}$ is the Boltzmann constant).
From these parameters we compute that about 50 layers contain Mott insulating cores, each of which consists of up to 2000 atoms. 
Deviations from theory are likely to arise because of incomplete thermalization. 
The tunneling timescale is expected to be sufficiently fast for equilibration within layers (see below). 
Yet, the slow inter-layer tunneling when approaching the final configuration hinders the formation of a globally thermalized state.
A more detailed analysis would require a full non-equilibrium model of coupled 2D layers.

A characteristic feature of a Mott insulator is a gapped excitation spectrum~\cite{Brinkman1970}, which we probe by recording $D$ in response to modulating the lattice depth at different frequencies $\nu$~\cite{Jordens2008}.
After loading the gas into the lattice, we sinusoidally modulate $V_Y$ for $40\,\mathrm{ms}$ by $\pm 10\%$. 
As $V_Y$ interferes with $V_X$ this leads to a modulation in tunneling $t_x$ ($t_y$) of $\sim \pm 7\%$ ($\mp 17\%$) as well as an additional modulation of $U$ by $\pm 3\%$ caused by the changing width of the Wannier functions.
For the whole parameter range the response of the system is within the linear regime of double occupancy creation \cite{greif_probing_2011}, where the creation rate is proportional to the energy absorption rate~\cite{Akiyuki2012}. 
In Fig.~\ref{fig2}(b), we plot both the response and the measured base level without modulation (arrows) for the same interactions as used in panel (a) and $N=80(2)\times 10^3$.
For weak interactions there is almost no detectable response.
When entering the Mott insulating regime we observe a gapped spectrum with a pronounced peak at $\nu=U/h$, corresponding to the creation of localized double occupancies. 

In Fig.~\ref{fig2}(c), we compare the peak position at $\nu=U/h$ obtained from gaussian fits to modulation spectra for various scattering lengths~\cite{supplementary} with the on-site interaction energy calculated using Wannier functions.
For weak interactions the \emph{ab initio} calculation of the Hubbard parameter $U$ agrees well with the measured value (see also Fig.~\ref{fig4}(d)).
Deviations are observed for the strongest interactions.
We attribute this effect to the deep optical lattice in one direction leading to a size of the Wannier function comparable to the scattering length and possibly higher band effects. 
A more detailed theory would however be necessary for a quantitative comparison in this regime. 

\begin{figure}[b]
    \includegraphics{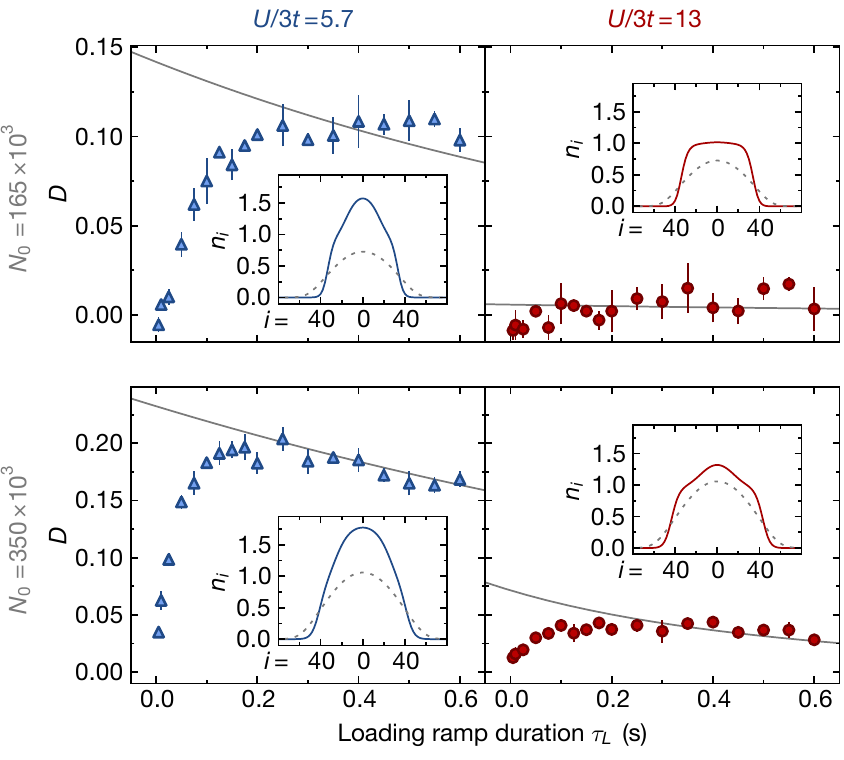}
    \caption{The lattice loading process. The panels show $D$ after loading ramps with varying duration $\tau_L$ for two interactions and two initial atom numbers. 
    The solid line is the expected $D$ from the high-temperature series expansion taking atom loss and heating during lattice loading into account~\cite{supplementary}.    
    The insets show the calculated equilibrium density profiles for the atomic cloud in the optical dipole trap (dashed) and in the lattice (solid lines), illustrating the required density redistribution during the loading. 
    Here the initial atom number and entropies before loading into the lattice were used.
    Errorbars in $D$ show the standard deviation of 3 measurements.
	}\label{fig3}
\end{figure}

The equilibration within the 2D honeycomb layers requires a change of the quantum many-body state during the lattice loading process. 
This is determined by the time necessary for the global density redistribution and the formation of correlations associated to the change in external potential. 
So far, equilibration dynamics have been investigated experimentally for bosonic atoms in optical lattices \cite{Gericke2007,Hung2010,Bakr2010}, whereas for strongly correlated fermions, the time evolution from the continuum to the Hubbard regime has not been studied yet.
In Fig.~\ref{fig3} we study the lattice loading process by measuring the resulting $D$ after an S-shaped intensity ramp \cite{supplementary} lasting between $\tau_L=5\,\mathrm{ms}$ and $\tau_L=600\,\mathrm{ms}$.
Both for intermediate ($a=242(1)\,a_0$) and strong interactions ($a=632(12) \,a_0$) we observe a fast rise of $D$ within roughly $200\,\mathrm{ms}$ followed by a slow decay.
We additionally plot the expected $D$ as derived from the high temperature series expansion (solid line) assuming global thermal equilibrium and taking into account atom loss and an independently determined heating rate \cite{supplementary}.
For $\tau_L \gtrsim 200 \, \mathrm{ms}$ the measured double occupancy agrees with the theoretical model. 
When comparing this timescale with the nearest-neighbor tunneling time of 6 ms in the honeycomb layers, this suggests that 200 ms is sufficient for density redistribution within the 2D layers (for the case of coupled layers, similar timescales are observed \cite{supplementary}).
The calculated density profiles for different interactions and atom numbers (insets Fig.~\ref{fig3}) indicate that the core density has to increase when loading the atoms from the dipole trap into the lattice. 
For very short ramp times this density redistribution cannot occur leading to densities in the trap center, which are too low. This is confirmed by the observed low values of $D$ as compared to theory for small~$\tau_L$.

\begin{figure}[bt]
    \includegraphics{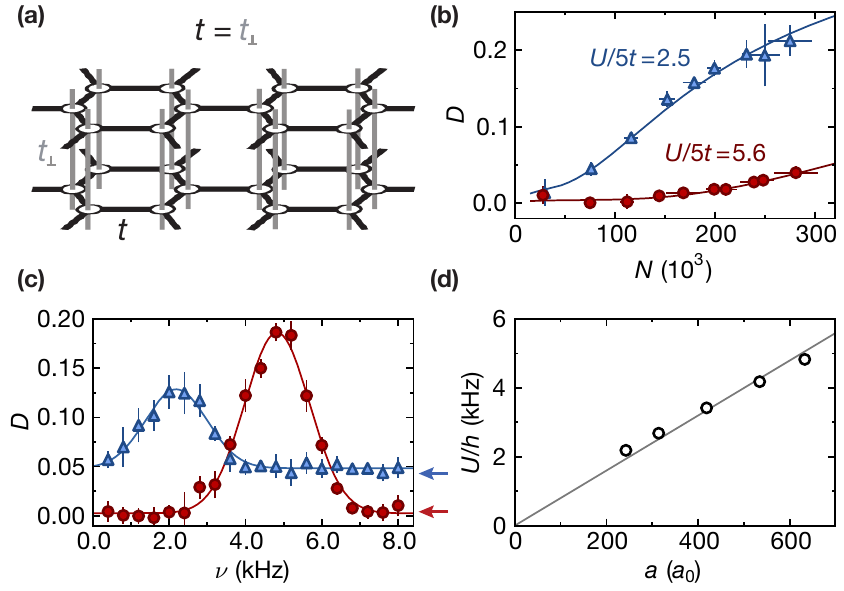}
    \caption{Coupled layers of artificial graphene.
    {\bf (a)} Detail of the coupled layer structure with $t=t_\perp$. The atoms populate about $80$ layers.
    {\bf (b)} The double occupancy $D$ versus atom number $N$ in the metallic and Mott insulating regime.
    Solid lines are theory predictions based on a high temperature series expansion.
    {\bf (c)} Excitation spectra for the interactions used in panel (b).
    The solid lines are gaussian fits to the spectra.
    Arrows show the reference value without modulation. 
    {\bf (d)} The on-site interaction energy $U$ compared to the theoretical expectation.
    Errorbars as in Fig.~\ref{fig2}.
    Data for additional interactions can be found in~\cite{supplementary}.
	}\label{fig4}
\end{figure}

The coupling between 2D layers is known to alter their physical properties as compared to mono-layer systems. 
For the case of real graphene, this has been used to modify the dispersion relation around the Dirac points \cite{Novoselov2006}.
In our experiment coupled honeycomb layers stacked as shown in Fig.~\ref{fig4}(a) can be produced, opening the possibility to simulate multi-layer systems with tunable interactions.
The tunneling between sites of adjacent layers~$t_{\perp}$ can be controlled via the lattice depth~$V_{\widetilde{Z}}$.
In the following we set $V_{\widetilde{Z}}=7E_R$ (corresponding to $t=t_{\perp}$) and investigate the dependence of double occupancy on atom number, see Fig.~\ref{fig4}(b).
The scattering length is set to the same values as in Fig.~\ref{fig3} and $\omega_{x,y,z}/2\pi = [55.7(7),106(1),57(1)]\,\mathrm{Hz}$. 
For weak repulsive interactions ($U/5t=2.5(3)$ with $U/h=2.18(4)\,\mathrm{kHz}$) the system is metallic, whereas for large interactions ($U/5t=5.6(7)$ with $U/h=4.82(2)\,\mathrm{kHz}$) the half-filled system is in the Mott insulating regime, signaled by a strong suppression of $D$.
We find excellent agreement with the theoretical predictions of the high-temperature series expansion using a connectivity of $5$.
The fitted entropy per particle is $s=1.8 k_{\mathrm{B}}$ for both interactions. As compared to the 2D measurements, we find only negligible deviations from the calculated double occupancy for the whole range of interactions \cite{supplementary}.
We attribute this to the fast tunneling time between layers leading to equilibration even between the honeycomb planes.

Both the uncoupled- and coupled-layer systems show a crossover from the metallic to the Mott insulating regime, however quantitative differences are observed in the double occupancy dependence for the case of coupled layers.
These differences originate in the altered lattice structure, which changes both the lattice connectivity and on-site interaction $U$. 
Using the same method as for the 2D data, we measure the lattice modulation spectra and find a reduction by about $25\%$ for the value of $U$ at the same scattering length, see Fig.~\ref{fig4}(c). 
For strong interactions a gapped excitation spectrum is found, as expected for a Mott insulating state.
The experimentally determined~$U$ is shown in Fig.~\ref{fig4}(d). 
In contrast to the 2D measurements, it does not deviate from the results obtained from lowest-band Wannier function overlap integrals even for the largest scattering lengths, owing to the weaker lattice depth along the coupled layer direction. 

In conclusion, we have investigated the properties of an artificial graphene system as a function of interactions. 
Mapping to a microscopic theory has provided insight into equilibration dynamics and the effect of coupling layers. 
The realization of a two-dimensional fermionic Mott insulator provides a platform for studying further strongly correlated phases, which have attracted particular interest in the honeycomb geometry, where spin-liquid and superconducting phases have been predicted~\cite{meng_quantum_2010, zhao_bcs-bec_2006, nandkishore_chiral_2012}.

\begin{acknowledgments}
We would like to thank Hari Manoharan and Leticia Tarruell for insightful discussions. We acknowledge SNF, NCCR-QSIT, and SQMS (ERC advanced grant) for funding. U. B. and W. H. acknowledge support from SFB-TR/49 and Forschergruppe FOR 801 by the DFG.
\end{acknowledgments}

\newpage

\makeatletter
\setcounter{section}{0}
\setcounter{subsection}{0}
\setcounter{figure}{0}
\setcounter{equation}{0}
\renewcommand{\bibnumfmt}[1]{[S#1]}
\renewcommand{\citenumfont}[1]{S#1}
\renewcommand{\thefigure}{S\@arabic\c@figure}
\renewcommand{\theequation}{S\@arabic\c@equation}
\makeatother

\onecolumngrid
  
\section{Supplemental Material}  
  
\subsection{Additional experimental data for isolated layers}

\begin{figure}[!ht] 
    \includegraphics{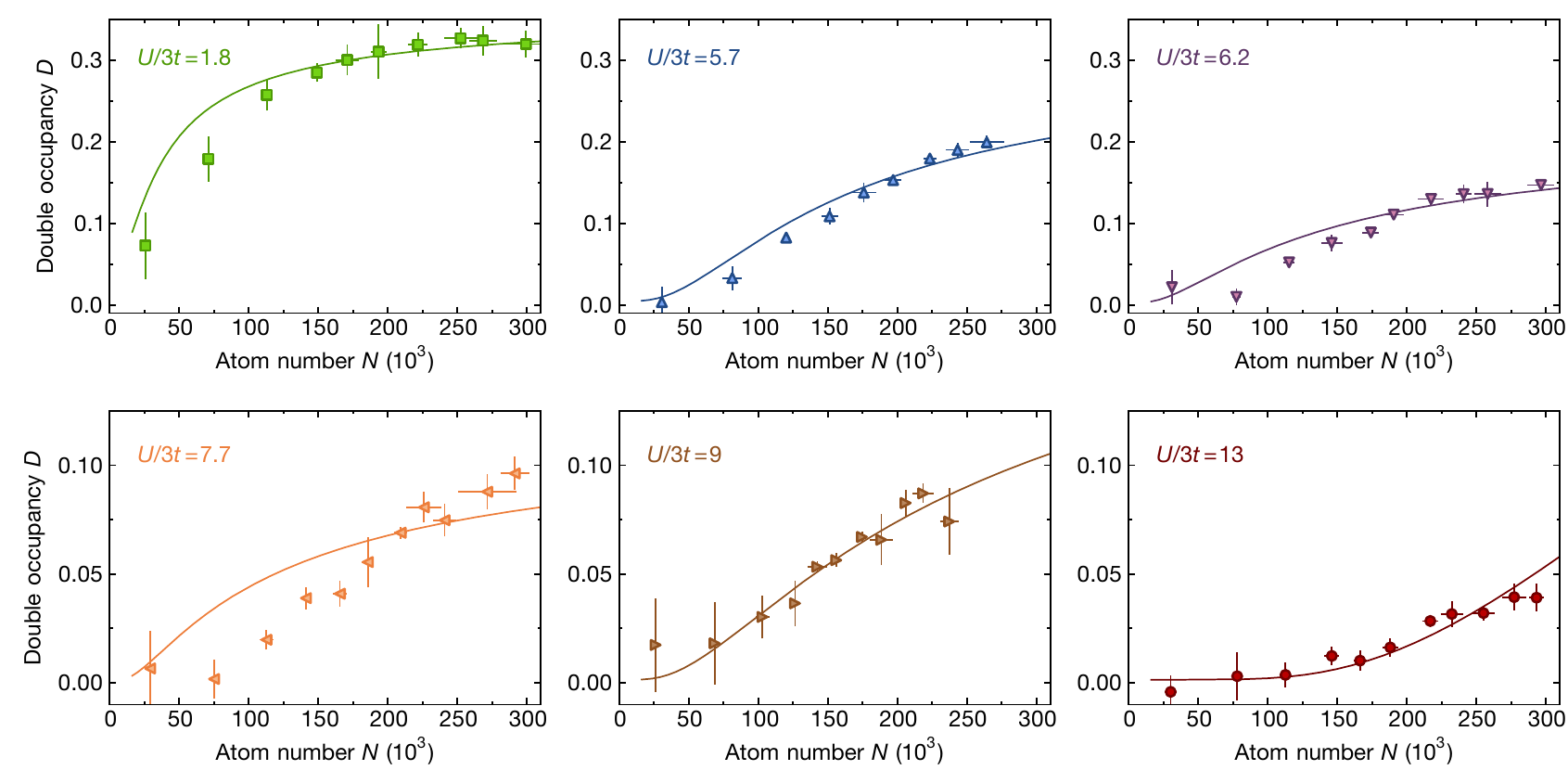}
    \caption{Observing the metal to Mott insulator crossover in artificial graphene. 
    The measured double occupancy~$D$ versus atom number $N$ is shown for additional interactions as compared to Fig.~2 in the main text with scattering lengths $a=[86(1),242(1),270(1),347(3),429(4),632(11)] a_0$.
    Solid lines are theoretical predictions from the high-temperature series expansion up to second order with fitted entropies of $s=[2.1,2.2,2.7,3.4,2.7,1.7]k_\mathrm{B}$ for $U/3t=[1.8(3),5.7(7),6.2(7),7.7(9),13(1)]$ respectively.
    Errorbars in $D$ and $N$ show the standard deviation of 5 measurements. 
	}\label{sfig_2D_DvsN}
\end{figure}

\begin{figure}[!ht]
    \includegraphics{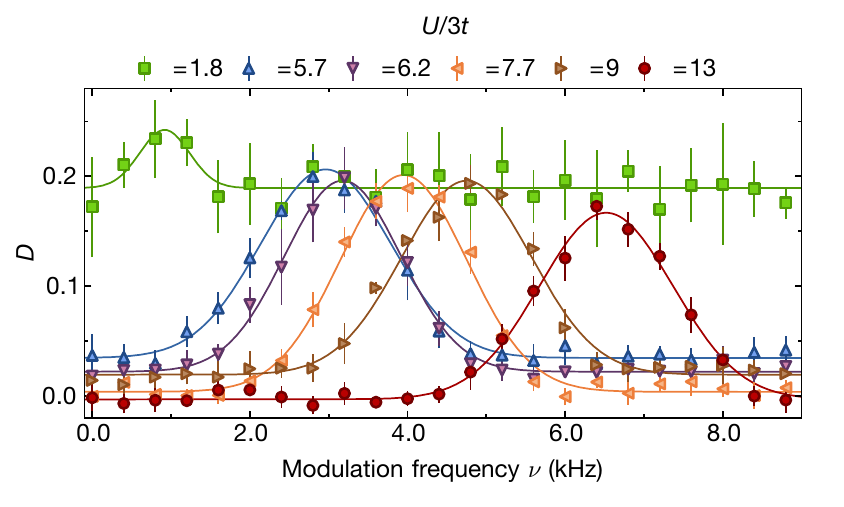}
    \caption{Excitation spectra obtained in the artificial graphene system by measuring $D$ after sinusoidal modulation of the lattice depth $V_Y$ for the interaction strengths used in Fig.~\ref{sfig_2D_DvsN}.
    The solid lines are gaussian fits to the spectra. 
    Errorbars as in Fig.~\ref{sfig_2D_DvsN}.
	}\label{sfig_2D_modulation}
\end{figure}

In this section we present experimental data supplementing Fig.~2 in the main text. 
Measurements of the double occupancy $D$ versus atom number $N$ for variable interactions are shown in Fig.~\ref{sfig_2D_DvsN}, together with the data already presented in the main manuscript. 
The excitation spectra for all data points in Fig.~2(c) are shown in Fig.~\ref{sfig_2D_modulation}. 
The scattering lengths used for this data are the same as for the measurements in Fig.~\ref{sfig_2D_DvsN}.

\subsection{Additional experimental data for coupled layers}

\begin{figure}[!ht]
    \includegraphics{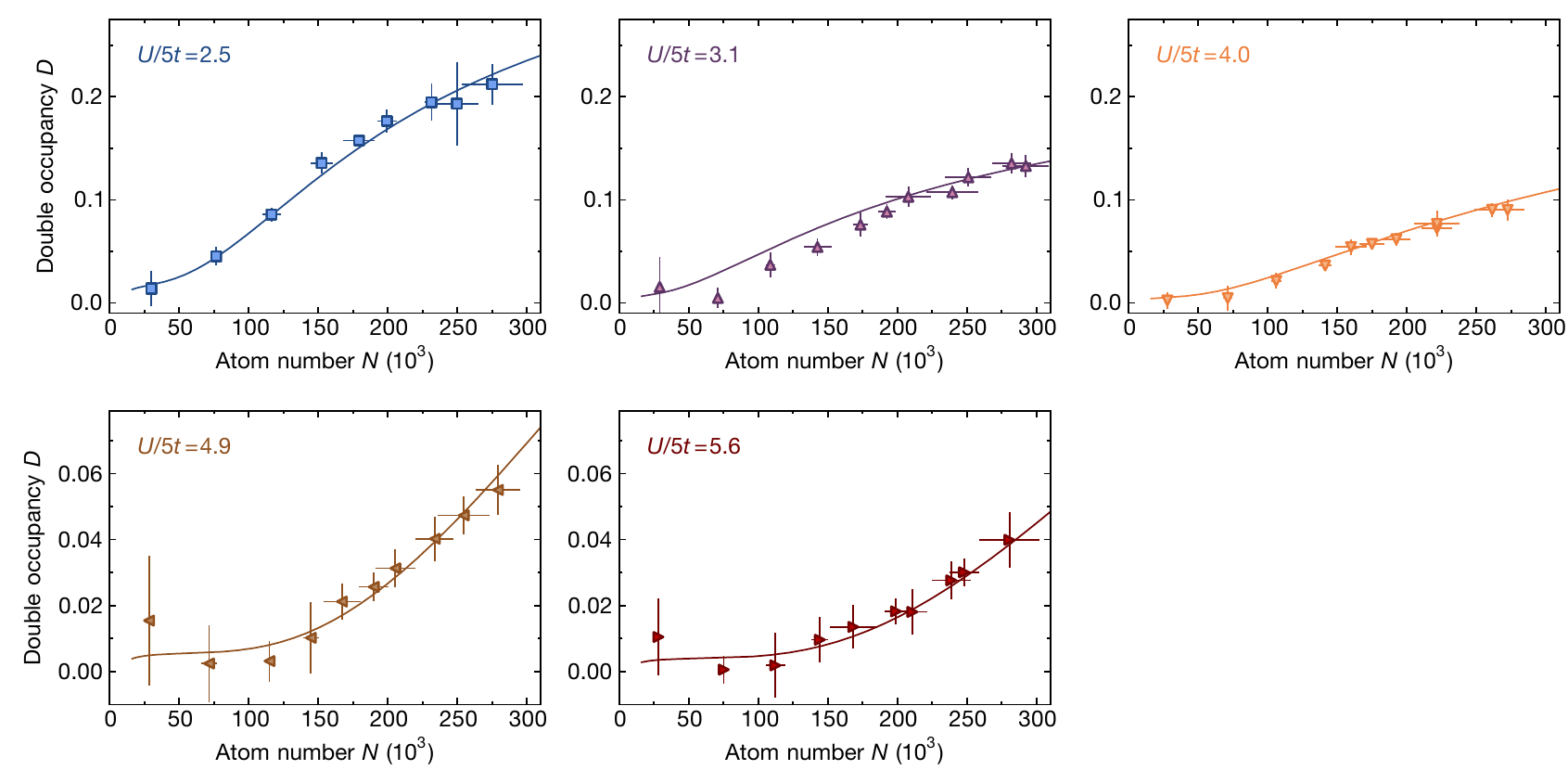}
    \caption{Observing the metal to Mott insulator crossover in coupled honeycomb layers. 
    The measured double occupancy~$D$ versus atom number $N$ is shown for additional interactions as compared to Fig.~4 in the main text. 
    The same scattering lengths as for the 2D measurements shown in Fig.~\ref{sfig_2D_DvsN} are used, except for the lowest scattering length.
    Solid lines are theoretical predictions from a high-temperature series expansion up to second order, from which we obtain fitted entropies of $s=[1.8,2.5,2.4,1.7,1.8]k_\mathrm{B}$ for $U/5t=[2.5(3),3.1(4),4.0(5),4.9(6),5.6(7)]$ respectively.
    Errorbars as in Fig.~\ref{sfig_2D_DvsN}.
	}\label{sfig_3D_DvsN}
\end{figure}

\begin{figure}[!ht]
    \includegraphics{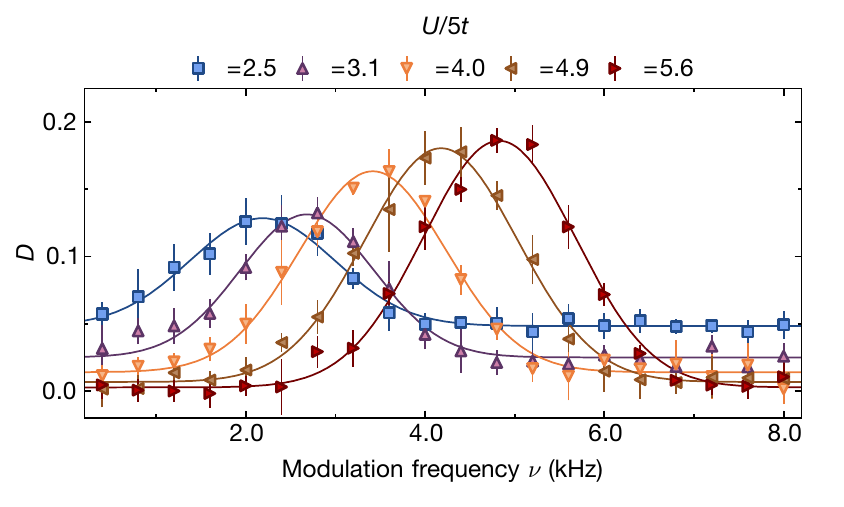}
    \caption{Excitation spectrum obtained in coupled honeycomb layers by measuring $D$ after sinusoidal modulation of the lattice depth $V_Y$ at the interaction strengths used in Fig.~\ref{sfig_3D_DvsN}.
    The solid lines are gaussian fits to the spectra.
    Errorbars as in Fig.~\ref{sfig_2D_DvsN}.
	}\label{sfig_3D_modulation}
\end{figure}

\begin{figure}[!ht]
    \includegraphics{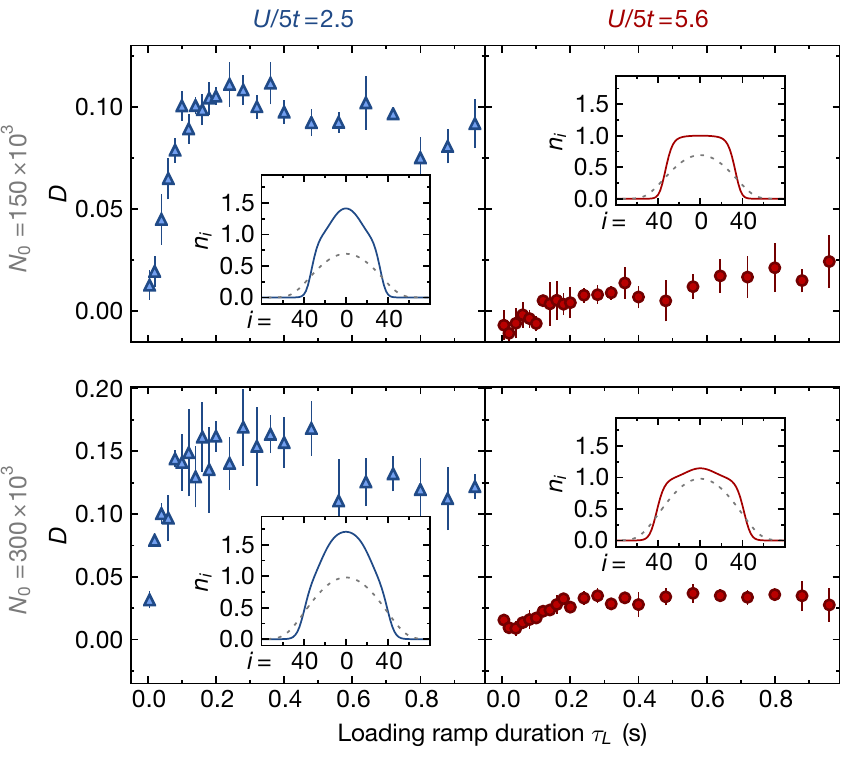}
    \caption{The lattice loading process for coupled honeycomb layers. The panels show $D$ after loading ramps with varying duration $\tau_L$ for two interactions and two initial atom numbers. 
    The insets show the calculated equilibrium density profiles for the atomic cloud in the optical dipole trap (dashed) and in the lattice (solid lines), illustrating the required density redistribution during the loading.
    Errorbars in $D$ show the standard deviation of 3 measurements.
	}\label{sfig_3D_thermalization}
\end{figure}

Data for the case of coupled honeycomb layers with $t_\perp=t$ supplementing Fig.~4 in the main text is shown in this section. Measurements of $D$ vs. $N$ for additional intermediate interactions can be found in Fig.~\ref{sfig_3D_DvsN}.
The corresponding excitation spectra, from which the extracted peak positions are shown in Fig.~4(d) of the main text, are given in Fig.~\ref{sfig_3D_modulation}. The scattering lengths are the same as for the data in Fig.~\ref{sfig_3D_DvsN}. We also studied the lattice loading process in coupled honeycomb layer systems by measuring $D$ for various lattice loading ramp durations $\tau_L$, see Fig.~\ref{sfig_3D_thermalization}. We conclude that the system is thermalized for lattice loading ramps with $\tau_L \gtrsim 200\,\mathrm{ms}$.

\subsection{Lattice loading}

For the loading of the lattice from the dipole trap we use an S-shaped intensity ramp to the final lattice depth $V_0$ with a total ramp time $\tau_L$. The full time-dependent expression of the intensity ramp is
\begin{equation}
V(\tau) = 3V_0\left(\frac{\tau}{\tau_L}\right)^3-2V_0\left(\frac{\tau}{\tau_L}\right)^2.
\end{equation}

The heating during lattice loading for variable times $\tau_L$ is measured by reversing the loading procedure and extracting the resulting entropy in the dipole trap. For ramp durations larger than $200$~ms we find a roughly linear increase in entropy with time. The theoretical predictions for the double occupancy versus loading time (solid lines in Fig. 3) are then obtained using the parameters from the double occupancy measurements shown in Figs.~\ref{sfig_2D_DvsN} and \ref{sfig_2D_modulation}.

\subsection{Wannier function calculation}

To date, the numerical calculation of Wannier functions, both in the solid state, as well as in the optical lattice community, has relied on the well established Marzari-Vanderbilt scheme \cite{Marzari1997_2,Souza2001_2}, worked out in great detail and implemented into a numerical package \cite{Mostofi2008685_2}. 
There, the Wannier states are determined by numerically minimizing the spatial variance as a function of a set of $d$ variables parametrizing the unitary transformation into the Wannier basis. 
For complex lattice structures this entails a minimization in a $d$-dimensional parameter space with the spatial spread function featuring local minima, requiring lattice-specific modifications \cite{Ibanez-Azpiroz2013-2_2}.
Here, we describe in more detail our alternative numerical method based on \cite{Kivelson1982_2} to calculate Wannier states, reducing the problem to the diagonalization of a $d\times d$ matrix. 
This intrinsically avoids the problem of local minima and is numerically highly efficient, scaling favorably with increasing lattice complexity. 

For clarity, we first describe the procedure for a one-dimensional system with a single potential minimum per lattice unit cell. 
We start by assuming that a set of Bloch states is given, each in the form
\spl{
\ket{k,\alpha}=\sum_n c_n^{(k,\alpha)} \ket{p=k+2 \pi n / a},
}
with the normalization condition $\sum_{n} |c_{n}^{(k,\alpha)}|^2=1$,  $\ket p$ being a pure momentum state and $a$ the lattice spacing. 
This corresponds to each real-space Bloch function being normalized in each unit cell. 
In terms of these, the projection operator onto a band $\alpha$ can be written as
\spl{
\mathcal P_{\alpha}=\sum_{k} \ket{k,\alpha} \bra{k,\alpha},
}
which is independent of the Bloch states' (indetermined and arbitrary) complex phases.

The central idea developed by Kivelson \cite{Kivelson1982_2} in 1982 is to consider the Wannier states as eigenstates of the operator~$\hat x_{\alpha}$
\spl{\
\hat x_{\alpha}=\mathcal P_{\alpha}\, \hat x \,\mathcal P_{\alpha}.
}
In many standard cases this definition coincides with the usual definition via the Fourier transform of Bloch states with the additional requirement of minimizing the spatial variance. 
We find that this definition is highly suitable for explicit calculation of the Wannier states by a numerical diagonalization of $\hat x_{\alpha}$. 
Within each band $\alpha$, the natural basis for the explicit representation of this operator is the Bloch basis, where the matrix elements can be expressed as the real-space integrals of terms involving the Bloch functions ${\psi}_{k,\alpha}(x)=\braket{x}{k,\alpha}$ over the entire spatial region of the lattice consisting of $L$ sites
\begin{equation}
	X_{k,k'}^{(\alpha)}=\int_{-\frac a 2}^{(L-\frac 1 2)a} {\psi}_{k,\alpha} ^*(x) \; {\psi}_{k',\alpha}(x) \; x \, dx.
\end{equation}

The real-space integration can be performed analytically and we obtain
\spl{
\label{Eq:Xkk_Operator_1D}
X_{k,k'}^{(\alpha)}&= \delta_{k,k'}   \frac {a(L-1)}{2} + a \,e^{i \frac{a}{2}(k-k')}
\sum_{n,n'=-\infty}^\infty (1-\delta_{k,k'}\,\delta_{n,n'})
    \frac{ (-1)^{n-n'}  \,  {c_{n}^{(k,\alpha)}}^*  c_{n'}^{(k',\alpha)}}{2\pi i(n-n')+ia(k-k')},
}
reducing the calculation of each matrix element to a numerically efficient summation. 
Diagonalizing the resulting matrix $X$ directly leads to the Wannier states (up to a complex phase) without any ambiguity.

Both eigenvalues (corresponding to the position of the respective Wannier state) and eigenstates at the edge of a finite system contain finite size effects. 
However, these decay exponentially towards the center. 
In fact, it is sufficient to determine one Wannier function per sublattice (i.e. two Wannier states for our bipartite honeycomb lattice) to obtain the entire basis set of orthogonal Wannier states. 
All other Wannier states are related and can directly be obtained from simple phase rotations of the eigenvector elements, as follows from the Wannier states being related to the Bloch states by a discrete Fourier transformation. 
It is thus useful to determine a Wannier state at the center of the lattice to minimize finite size effects.

\subsubsection{Two-dimensional honeycomb lattice}

\begin{figure}[h!]
    \includegraphics[width=0.95\columnwidth]{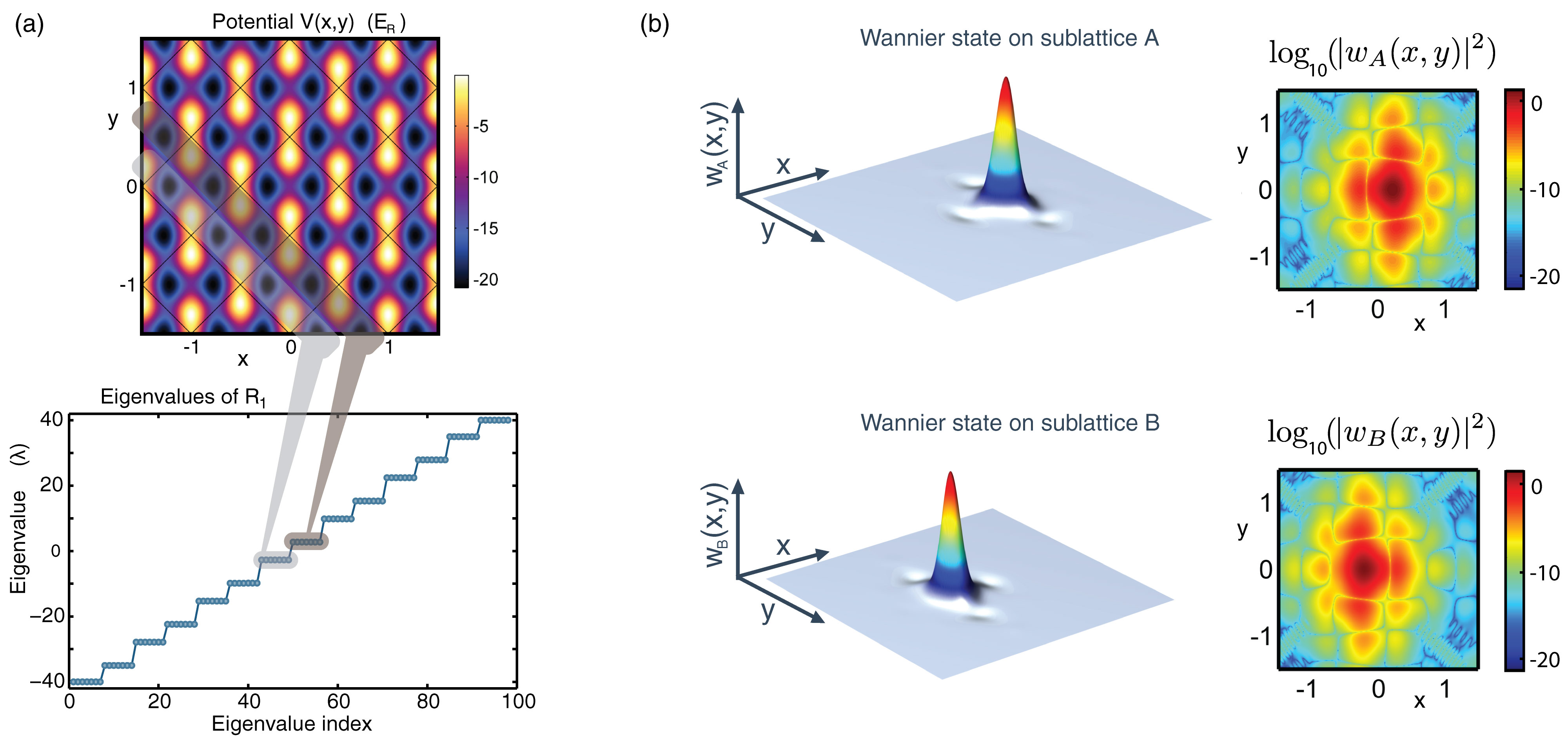}
    \caption{\label{FIG:R1_spectrum}
    {\bf (a)} Spectrum of the operator $R_1$ for a lattice consisting of L = 7 units cells per dimension, corresponding to 98 physical sites. 
    Each degenerate set of eigenvalues corresponds to a subset of states maximally localized along the direction $\mathbf{b}_1$, but not necessarily localized along $\mathbf{b}_2$, as indicated by the respective shaded regions.
    {\bf (b)} The two-dimensional Wannier functions in each plane of the honeycomb lattice obtained from our numerical procedure on both sublattices for the parameters chosen in the experiment $[V_{\overline X}, V_X, V_Y]/E_R = [14, \, 0.8, \, 6.45]$ and $\theta=\pi$. 
    Here the Wannier function length scale is normalized to $\lambda$.
    }
\end{figure}

In the two-dimensional case, the Bloch state with quasi-momentum $\bk$ in band $\alpha$ is of the form
\spl{
\ket{\bk,\alpha}=\sum_{n_1,n_2} c_{n_1,n_2}^{(\bk,\alpha)} \ket{\bp= \bk + n_1 \bb_1 + n_2 \bb_2}
}
and is similarly obtained from a common band structure calculation. For lattice geometries such as the honeycomb lattice with two or more potential minima per lattice unit cell, one has to allow for maximally localized Wannier states to be composed of Bloch states from multiple energy bands. We therefore define the projection operator onto a suitable subset of bands $\mathcal A$ as
\spl{
\mathcal P_{\mathcal A}=\sum_{\alpha\in \mathcal A, \, \bk} \ket{\bk,\alpha} \bra{\bk,\alpha}
}
and consider the Wannier states to be eigenstates of suitable position operators projected onto $\mathcal A$. Generally, in the higher-dimensional case, the Wannier states are maximally localized along the directions of the reciprocal lattice vectors, which are ($k_L=2\pi/\lambda$)
\begin{equation}
\bb_1=k_L (\be_x + \be_y) \qquad \bb_2=k_L (\be_x - \be_y)
\end{equation}
for our honeycomb lattice. 
We therefore define the real-space coordinate operators along these directions
\spl{
\hat r_1&=\bb_1 \cdot \vvec{\hat x}{\hat y}=k_L(\hat x + \hat y)\\
\hat r_2&=\bb_2 \cdot \vvec{\hat x}{\hat y}=k_L(\hat x - \hat y),
}
and the two-dimensional Wannier states are simultaneous eigenstates of both band-projected operators $R_1=\mathcal P_{\mathcal A}\, \hat r_1 \,\mathcal P_{\mathcal A}$ and $R_2=\mathcal P_{\mathcal A}\, \hat r_2 \,\mathcal P_{\mathcal A}$. 
Parametrizing the quasi-momentum by $\bk = \frac{m_1}{L} \bb_1+ \frac{m_2}{L} \bb_2$ with integer $m_1$ and $m_2$ for a two-dimensional lattice with $L$ lattice sites along each dimension and defining the collective index function $I(m_1,m_2,\alpha)$, which maps every Bloch state onto a unique integer value and noting that $\mathcal P_{\mathcal A} \ket{k,\alpha}=\ket{k,\alpha}$ if $\alpha \in {\mathcal A}$ , the matrix elements of the band-projected position operators determined from the real-space integration are
\spl{
R_{I(m_1,m_2,\alpha),I(m_1',m_2',\alpha')}^{(1,2)}&=\bra{\bk,\alpha} k_L(\hat x \pm \hat y) \ket{\bk',\alpha'}\\
&=k_L \int_{\mbox{\tiny{r.s.l.}}} d^2r \; \sum_{\stackrel{n_1,n_2}{n_1',n_2'}}  {c_{n_1,n_2}^{(\bk, \alpha)}}^* \;  c_{n_1',n_2'}^{(\bk', \alpha')} 
e^{-i (n_1 \, \bb_1 +n_2 \, \bb_2+\bk) \cdot \br} \, e^{i (n_1' \, \bb_1 +n_2' \, \bb_2+\bk') \cdot \br} \; (x\pm y),
}
where the integration is to be performed over the entire real-space lattice ($\mbox{r.s.l.}$). 
The explicit real-space integration can be performed analytically in full analogy to the 1D case, leading to a similar expression involving only the summation over $n_1$ and $n_2$. 
To determine the Wannier states as the simultaneous eigenstates, one can first diagonalize $R^{(1)}$, for which a typical spectrum is shown in Fig.~\ref{FIG:R1_spectrum}(a).

The spectrum of $R_1$ is composed of degenerate plateaus of eigenvalues \footnote{Up to deviations from finite size effects, which decay rapidly and are exponentially suppressed for states in the bulk of the lattice.}, each corresponding to a subspace of states maximally localized along $\bb_1$, but with arbitrary localization properties along $\bb_2$. 
To obtain the final Wannier states, the operator $R_2$ is diagonalized within one such degenerate subspace. 
A typical Wannier state obtained in this manner for the honeycomb lattice considered in this work is shown in Fig.~\ref{FIG:R1_spectrum}(b).

\subsubsection{Realness of Wannier functions}

A property which has been much discussed but not resolved is why the real-space representation of the Wannier functions obtained from the Marzari-Vanderbilt minimization approach is real (up to an arbitrary constant complex phase factor) if the spatial variance is minimal. 
This property follows naturally within our approach for lattice Hamiltonians, which are invariant under time reversal symmetry: in this case, the real-space wave functions of all energy eigenstates can be chosen purely real. 
This seems to contradict the complex form of the Bloch functions for a system which is infinitely large or has periodic boundary conditions, but is easily resolved by noting that the Bloch states $\ket{k,\alpha}$ and $\ket{-k,\alpha}$ are pairwise degenerate. 
By performing a unitary transformation within each such two-dimensional subspace to an equivalent basis $(\ket{k,\alpha}+\ket{-k,\alpha})/\sqrt 2$ and $(\ket{k,\alpha}-\ket{-k,\alpha})/\sqrt 2$ for some suitable choice of initial phases, states with purely real-space wave functions can be formed. 
Clearly, the definition via eigenstates is basis-independent and can equally well be performed in this alternative purely real basis without altering the resulting Wannier states. 
However, it is directly evident that the matrix elements of the operator $\hat x_{\mathcal{A}}$ are purely real in this basis, since they can be expressed as integrals of a product of three real functions (two real energy eigenfunctions and the position $x$). 
Hence, the representation of $\hat x_{\mathcal{A}}$ in this basis is a real, symmetric matrix. 
Since the eigenvector elements of a symmetric matrix can be chosen to be purely real and the Wannier states can be written as superpositions of these elements and the corresponding real basis functions, the Wannier functions are purely real.

\end{document}